\documentclass[10pt]{elsart}

\begin{document}

\begin{frontmatter}

\title{\bf{On the Reduction of CP Violation Phases}}

\author{J.I. Silva-Marcos\thanksref{juca}}
\address{Instituto Superior T\'ecnico, Grupo de F{\'\i}sica de 
Part{\'\i}culas, Av. Rovisco Pais, 1049-001 Lisboa, Portugal}

\thanks[juca]{E-mail: Joaquim.Silva-Marcos@cern.ch}

\begin{abstract}
We describe a method that is used to reduce, significantly, the number of CP
violating complex phases in the Yukawa parameters. With this Reduction of
Complex Phases (RCP) we obtain only one CP violating complex phase in the
case where the neutrinos have an (effective) $3\times 3$ Majorana mass matrix. For the
See Saw extension of the SM with three righthanded neutrinos, and in
connection with CP violation in leptogenis, we reduce the usual 6 complex
phases to only 2.
\end{abstract}
\end{frontmatter}

Understanding the origin of CP violation is one of the outstanding problems
of particle physics. In many extensions of the Standard Model (SM), e.g. in
the See Saw mechanism and leptogenisis theories in neutrino physics \cite
{yanagida} \cite{many} or in supersymmetric theories \cite{dent}, one has a
large number of complex Yukawa coupling parameters which lead to CP
violation \cite{also}. However, a large number of (complex) parameters may
also obscure some fundamental structure leading to CP violation. In
contrast, a specific choice of parameters used to describe CP violation may
play an important and crucial r\^ole in solving the mechanisms involved.

In this letter, we describe a simple method that can reduce, significantly,
the number of CP violating complex phases in the Yukawa parameters, and
which can be applied to many extensions of the SM and theories. Here, the
Reduction of Complex Phases (RCP) is used to minimize the number of complex
phases in neutrino physics. In particular, we reduce (by a factor of 3) the
number of CP violating complex phases in the case where the neutrinos have
an (effective) $3\times 3$ Majorana mass matrix, and in the case of the See
Saw extension of the SM with three righthanded neutrinos. We also study
briefly some implications of RCP to CP violation invariants relevant for
leptogenis.

{\bf Reduction of Complex Phases}

RCP is based on the following, rather simple, transformation: it is always
possible to rotate any pair of complex parameters $(a+ix,b+iy)$ with an
orthogonal matrix such that one (say the first parameter) becomes real, i.e. 
\begin{equation}
\label{complex}\left( 
\begin{array}{c}
a+ix \\ 
b+iy 
\end{array}
\right) \longrightarrow \left( 
\begin{array}{c}
a^{\prime } \\ 
b^{\prime }+iy^{\prime } 
\end{array}
\right) =\left[ 
\begin{array}{cc}
\cos (\theta ) & -\sin (\theta ) \\ 
\sin (\theta ) & \cos (\theta ) 
\end{array}
\right] \cdot \left( 
\begin{array}{c}
a+ix \\ 
b+iy 
\end{array}
\right) 
\end{equation}
Clearly, the first parameter will have no imaginary part if one chooses $%
\tan (\theta )=x/y$.

Next, we prove that any Hermitian matrix $H$ can be written as 
\begin{equation}
\label{hermitian}H=\left( OK_\alpha \widehat{O}\right) ^{\dagger }\cdot
D\cdot OK_\alpha \widehat{O} 
\end{equation}
where $D$ is a real diagonal matrix, $O$, $\widehat{O}$ are orthogonal
matrices and $K_\alpha $ is a diagonal unitary matrix with only one complex
phase, $K_\alpha ={\rm {diag}\left[ 1,1,e^{i\alpha }\right] }$. Taking a
general Hermitian matrix 
\begin{equation}
\label{hermitian1}H=\left[ 
\begin{array}{lll}
r_1 & c_4 & c_6 \\ 
c_4^{*} & r_2 & c_5 \\ 
c_6^{*} & c_5^{*} & r_3 
\end{array}
\right] 
\end{equation}
where the $r_i$ are real and the $c_i$ are complex entries and rotating the
2-3 position with and orthogonal matrix $O_{23}$, i.e. 
\begin{equation}
\label{just1}
\begin{array}{c}
H\longrightarrow H^{\prime }=O_{23}^T\cdot H\cdot O_{23} 
\end{array}
\end{equation}
where%
$$
O_{23}=\left[ 
\begin{array}{lll}
1 & 0 & 0 \\ 
0 & \cos (\theta _{23}) & \sin (\theta _{23}) \\ 
0 & -\sin (\theta _{23}) & \cos (\theta _{23}) 
\end{array}
\right] 
$$
and applying the RCP method described in Eq. (\ref{complex}), one obtains $%
c_4\rightarrow c_4^{\prime }=r_4$ real. Then, by rotating the 1-2 position,
we find $c_6=r_6$ real, and subsequently by the rotating again the 2-3
position we can even have $r_6=0$. Thus, any Hermitian matrix can be written
in the following way: 
\begin{equation}
\label{justa}
\begin{array}{c}
H=\widehat{O}^T\cdot \left[ 
\begin{array}{lll}
r_1 & r_4 & 0 \\ 
r_4 & r_2 & r_5\ e^{i\alpha } \\ 
0 & r_5\ e^{-i\alpha } & r_3 
\end{array}
\right] \cdot \widehat{O} 
\end{array}
\end{equation}
where $\widehat{O}$ is some general (appropriate) orthogonal matrix.
Finally, one concludes: 
\begin{equation}
\label{just2}
\begin{array}{c}
H=\widehat{O}^T\cdot K_\alpha ^{*}\cdot \left[ 
\begin{array}{lll}
r_1 & r_4 & 0 \\ 
r_4 & r_2 & r_5 \\ 
0 & r_5 & r_3 
\end{array}
\right] \cdot K_\alpha \cdot \widehat{O}=\left( OK_\alpha \widehat{O}\right)
^{\dagger }\cdot D\cdot OK_\alpha \widehat{O} 
\end{array}
\end{equation}
where $O$ is just another orthogonal matrix, $D$ real diagonal and $K_\alpha
={\rm {diag}}$ ${\rm \left[ 1,1,e^{i\alpha }\right] }$. Furthermore, taking
into account that any orthogonal matrix results from three rotations $%
O=O_{12}O_{23}O_{12}^{\prime }$, and that any $O_{12}$ commutes with $%
K_\alpha $, one finds that the combination $OK_\alpha \widehat{O}$ is a
product of just 5 different rotations and the diagonal unitary matrix $%
K_\alpha $ with only one complex phase: 
\begin{equation}
\label{combi}OK_\alpha \widehat{O}=O_{12}O_{23}O_{12}^{\prime }\ K_\alpha \ 
\widehat{O}_{23}\widehat{O}_{12} 
\end{equation}
Thus, any Hermitian matrix is a combination of its 3 eigenvalues, 5
rotations and only one complex phase. This gives a total of 9 parameters,
which corresponds exactly with the same number of parameters as in the
general case for any Hermitian matrix (as in Eq. (\ref{hermitian1})), where
we have 6 real parameters and 3 complex phases.

We shall now apply the reduction of complex phases to several cases and
scenarios.

{\bf 1 - Majorana neutrinos}

Consider the case where we have (only) 3 lefthanded Majorana type neutrinos
with some Majorana mass matrix. This may also be an effective Majorana mass
matrix. Without loss of generality, one can find a weak-basis where the the
Majorana neutrino mass matrix is real and diagonal while the charged lepton
mass matrix $M_e$ is Hermitic. Then, applying RCP in this weak-basis, one
obtains 
\begin{equation}
\label{just}
\begin{array}{ccc}
M_\nu =D_\nu \equiv {\rm {diag}}\left[ {\rm m_{\nu _1},m_{\nu _2},m_{\nu _3}}%
\right] & ; & M_e=U^{\dagger }\cdot D\cdot U 
\end{array}
\end{equation}
where $U=OK\widehat{O}$, the combination described in Eq. (\ref{combi}), is
a unitary matrix with only one complex phase and five angle rotations: 
\begin{equation}
\label{unitary}
\begin{array}{ccc}
U=O_{12}O_{23}O_{12}^{\prime }\ K_\alpha \ \widehat{O}_{23}\widehat{O}_{12}
& ; & K_\alpha ={\rm {diag}}\left[ {\rm 1,1,e^{i\alpha }}\right] 
\end{array}
\end{equation}

$U$ corresponds, of course, to the neutrino mixing matrix which, for the
case of Majorana neutrinos, will have $3+1+2=6$ physical parameters: 3
angles, one CP violating phase of the CKM-type and 2 CP violating complex
phases of the Majorana type. In order to find these, we write $U$ also in
the following combination: 
\begin{equation}
\label{ckmtype}U=K_o\cdot V\cdot K_M
\end{equation}
where $K_o={\rm {diag}\left[ e^{i\phi _1},e^{i\phi _2},e^{i\phi _3}\right] }$%
, $K_M={\rm {diag}\left[ 1,e^{i\alpha _M},e^{i\beta _M}\right] }$ and $V$ is
a (CKM-type) unitary matrix where the first line and row are real. As can be
seen, by substituting this combination for $U$ in $M_e$ of Eq. (\ref{just}),
the diagonal unitary matrix $K_o$ is unphysical and does not contribute to $%
M_e$. The two physical unitary matrices are $V$ and $K_M$. The matrix $K_M=%
{\rm {diag}\left[ 1,e^{i\alpha _M},e^{i\beta _M},\right] }$ contains the two
Majorana type CP violating complex phases $(\alpha _M,\beta _M)$, and $V$
contains the CKM-type CP violating phase $\delta _{KM}$. All 3 phases $%
(\alpha _M,\beta _M,\delta _{KM})$ are functions of the initial 5 angles $%
(\theta _{12},\theta _{23},\theta _{12}^{\prime },\widehat{\theta }_{23},%
\widehat{\theta }_{12})$ contained in $O_{12}$, $O_{23}$, $O_{12}^{\prime }$%
, $\widehat{O}_{23}$, $\widehat{O}_{12}$, and the unique complex phase $%
\alpha $ contained in $K_\alpha $. However, comparing the two formulas for $U
$ in Eq. (\ref{unitary}) and Eq. (\ref{ckmtype}), it is quite clear that all
CP violation, be it of the CKM-type or the Majorana type, is controlled by
the one single complex phase $\alpha $ and that $(\alpha _M,\beta _M,\delta
_{KM})=0$ if $\alpha =0$. Furthermore\footnote{%
All angles and phases are given mod $\pi $.}, $\delta _{KM}=0$ if $\theta
_{23}=0$ or $\widehat{\theta }_{23}=0$. In this case, the matrix $K$
commutes with the remaining matrices $O_{12}$, $O_{12}^{\prime }$ or $%
\widehat{O}_{12}$. For the Majorana type CP violating phases, we obtain $%
(\alpha _M,\beta _M)=0$ if $\theta _{23}=0$ or if $\theta _{12}=0$. We find
also $(\alpha _M,\beta _M)=(0,\alpha )$ if $\widehat{\theta }_{23}=0$ (for $%
\theta _{12}$ and $\theta _{23}\neq 0$). Explicitly, one has 
\begin{equation}
\label{epli}
\begin{array}{l}
\tan (\delta _{KM})=\sin (\alpha )\ \sin (\theta _{23})\ \sin (
\widehat{\theta }_{23})\ \ A_1 \\ \tan (\alpha _M)=\sin (\alpha )\ \sin
(\theta _{12})\ \sin (\theta _{23})\ \sin (
\widehat{\theta }_{23})\ \ A_2 \\ \tan (\beta _M)=\tan (\alpha )\ \sin
(\theta _{12})\ \sin (\theta _{23})\ \ A_3
\end{array}
\end{equation}
where $A_1$, $A_2$ and $A_3$ are functions of sinuses and cosinuses of $%
(\alpha $, $\theta _{12}$, $\theta _{23}$, $\theta _{12}^{\prime }$, $%
\widehat{\theta }_{23}$, $\widehat{\theta }_{12})$; $A_1$ is well defined if 
$\alpha $ or $\theta _{23}$ or $\widehat{\theta }_{23}$ are zero, $A_2$ is
also well defined if $\alpha $ or $\theta _{12}$ or $\theta _{23}$ or $%
\widehat{\theta }_{23}$ are zero, and the same applies to $A_3$ if $\alpha $
or $\theta _{12}$ or $\theta _{23}$ are zero, furthermore, $A_3=1/\sin
(\theta _{12})\sin (\theta _{23})$ if $\widehat{\theta }_{23}=0$ (for $%
\theta _{12}$ and $\theta _{23}\neq 0$). A special case occurs for $\delta
_{KM}$ when $(\theta _{12},\theta _{12}^{\prime })=0$. It can be readily
verified from Eq. (\ref{unitary}) that, in this case, $U_{13}=0$ and that $%
A_1=0$; thus, $\delta _{KM}=0$.

{\bf 2 - See Saw}

Let us now consider the see saw scenario with 3 lefthanded and 3 right-
handed neutrinos, and where the neutrinos have only a Dirac mass matrix $M_D$
and a righthanded Majorana mass matrix $M_R$. As argued in Ref. \cite{branco}%
, one can always find a weak-basis where the righthanded Majorana neutrino
mass matrix and charged lepton mass matrix are diagonal and real, while the
Dirac neutrino mass matrix $M_D$ has 6 complex phases, i.e. 
\begin{equation}
\label{full}
\begin{array}{ccc}
M_\nu =\left( 
\begin{array}{cc}
0 & M_D \\ 
M_D^T & D_R 
\end{array}
\right) & ; & M_e=D_e 
\end{array}
\end{equation}
where $D_R\equiv {\rm {diag}\left[ M_1,M_2,M_3\right] =}$ $M_R$, $D_e\equiv 
{\rm {diag}\left[ m_e,m_\mu ,m_\tau \right] }$and the $(M_D)_{i1}$ are real.
Thus, we have a total of 3+3+9+6=21 parameters.

Next, we apply RCP to the see saw mechanism. To do this, one must first
consider a different weak-basis from the one in Eq. (\ref{full}).
Transforming (only) the lefthanded neutrinos with a unitary matrix $U$: 
\begin{equation}
\label{pli1}\nu _{L_i}\longrightarrow U_{ij}\ \nu _{L_j} 
\end{equation}
and keeping the charged lepton mass matrix Hermitic, Eq. (\ref{full})
becomes 
\begin{equation}
\label{full1}
\begin{array}{ccc}
M_\nu =\left( 
\begin{array}{cc}
0 & U^T\ M_D \\ 
M_D^T\ U & D_R 
\end{array}
\right) & ; & M_e=U^{\dagger }\ D_eU 
\end{array}
\end{equation}
This structure for the neutrinos as well as for the charged leptons has the
same physical content as the one in Eq. (\ref{full}). However, we may now
choose $U$ such that the new Dirac neutrino mass matrix $U^T\ M_D$ is also
Hermitic\footnote{%
This is clear. Every matrix $M_D$ can be diagonalized by a biunitary
transformation: $W\ M_D\ V=D$. Thus, $M_D=W^{\dagger }D\ V^{\dagger }$ and
we find that by multiplying $M_D$ on the left by a suitable unitary matrix $%
(VW)\ M_D=V\ D\ V^{\dagger }$, we obtain an Hermitic matrix.}. Thus, there
exists a weak-basis where both the Dirac neutrino and charged lepton mass
matrices $M_D$ and $M_e$ are Hermitic, while the (righthanded) Majorana
neutrino mass matrix $M_R=D_R$ is diagonal.

We are now ready to apply the reduction of complex phases, and may conclude,
as in Eq. (\ref{hermitian}), that there is a weakbasis where%
$$
\begin{array}{l}
M_D=\left( O_DK_D 
\widehat{O}_D\right) ^{\dagger }D\ O_DK_D\widehat{O}_D \\ M_e=\left( O_eK_e 
\widehat{O}_e\right) ^{\dagger }D_e\ O_eK_e\widehat{O}_e \\ M_R=D_R 
\end{array}
$$
where the $O$'s are orthogonal, the $D$'s are real and diagonal and the $K$%
's are unitary diagonal matrices with, each, only one complex phase. By
transforming again the lefthanded neutrinos, as in Eqs. (\ref{pli1}, \ref
{full1}), one can even find a weak-basis where $M_D$ is somewhat simplified: 
\begin{equation}
\label{only}
\begin{array}{l}
M_D=D\ O_DK_D 
\widehat{O}_D \\ M_e=\left( O_eK_e 
\widehat{O}_e\right) ^{\dagger }D_e\ O_eK_e\widehat{O}_e \\ M_R=D_R 
\end{array}
\end{equation}
and where the $O_e$'s and $K_e$'s have been redefined.

Thus, we find, in addition to the result obtained in \cite{branco}, a
weak-basis where the total number of complex phases in the see saw model is
(not 6) but only 2. However, it must be kept in mind that the total number
of parameters is still the same as in \cite{branco}, i.e. 3 masses, 5 angles
and 1 phase in $M_D$ and $M_e$, and 3 masses in $M_R$: $2\times (3+5+1)+3=21$%
.

With respect to CP violation, it is clear that all weak-basis invariants
measuring CP violation, dependent on the lepton mass matrices, will be zero
if both complex phases $\alpha _D,\alpha _e$ in $K_D$ and $K_e$ are zero.
E.g.

\begin{equation}
\label{invarianta}
\begin{array}{l}
I_\nu =Tr[M_D^{\dagger }M_D,M_R^{\dagger }M_R]^3=0\qquad if\qquad \alpha
_D=0 \\ 
I_{e\nu }=Tr[M_D^{*}M_D^T,M_eM_e^{\dagger }]^3=0\qquad if\qquad \alpha _e=0 
\end{array}
\end{equation}
Of course, $(I_\nu ,I_{e\nu })=0$ does not imply that only $(\alpha
_D,\alpha _e)=0$ . To study under which conditions $(I_\nu ,I_{e\nu })=0$,
it is useful to write $O_DK_D\widehat{O}_D$ in $M_D$ and $O_eK_e\widehat{O}%
_e $ in $M_e$ as in Eq. (\ref{combi}): 
\begin{equation}
\label{odoe}
\begin{array}{l}
O_DK_D 
\widehat{O}_D=O_{12}^DO_{23}^DO_{12}^{D\prime }\ K_D\ \hat O_{23}^D\hat
O_{12}^D \\ O_eK_e\widehat{O}_e=O_{12}^eO_{23}^eO_{12}^{e\prime }\ K_e\ \hat
O_{23}^e\hat O_{12}^e 
\end{array}
\end{equation}
Clearly, as explained in the previous paragraph we find also 
\begin{equation}
\label{invariant}
\begin{array}{l}
I_\nu =0~\ if~\ \alpha _D=0~\ or~\ \theta _{23}^D=0~\ or~\ 
\widehat{\theta }_{23}^D=0~\ or~\ ~(\theta _{12}^D,\theta _{12}^{D\prime
})=0 \\ I_{e\nu }=0~\ if~\ \alpha _e=0~\ or~\ \theta _{23}^e=0~\ or~\ 
\widehat{\theta }_{23}^e=0~\ or~\ ~(\theta _{12}^e,\theta _{12}^{e\prime
})=0 
\end{array}
\end{equation}

{\bf 3 - Leptogenisis}

Next we consider, as in Ref. \cite{branco}, another class of weak-basis
invariants which measure CP-violation and are also relevant for
leptogenisis: 
\begin{equation}
\label{week}
\begin{array}{l}
I_1=ImTr[h_DH_RM_R^{*}h_D^{*}M_R] \\ 
I_2=ImTr[h_DH_R^2M_R^{*}h_D^{*}M_R] \\ 
I_3=ImTr[h_DH_R^2M_R^{*}h_D^{*}M_RH_R] 
\end{array}
\end{equation}
where $h_D=M_D^{\dagger }M_D$ and $H_R=M_R^{\dagger }M_R$. Using the
parametrization defined in Eqs. (\ref{only}, \ref{odoe}), one obtains the
following relations for these weak-basis invariants 
\begin{equation}
\label{week1}I_i=\sin (\alpha _D)\ \sin (\theta _{23}^D)\ B_i 
\end{equation}
where the $B_i$ are polynomials in sinuses and cosinuses of $\theta _{ij}^D$%
, $\theta _{12}^{D\prime }$, $\widehat{\theta }_{ij}^D$ and $\alpha _D$ and
the eigenvalues of $M_D$ and $M_R$. Clearly, $I_i=0$ if $\alpha _D=0~\ $or$%
~\ \theta _{23}^D=0$.$~$

{\bf 4 - CP violation and low energy effective theory}.

Our parametrization is useful to study all kinds of conditions under which
CP-violation related to leptogenis will occur. E.g., it can be verified that
the invariants $I_i\neq 0$, (i.e. CP-violation will still occur) even when
all forms CP-violation are absent in the approximate effective low energy
theory with an effective neutrino mass matrix 
\begin{equation}
\label{eff}m_{eff}=-M_D\ M_R^{-1}\ M_D^T 
\end{equation}
Thus, the two effective low energy Majorana phases and CKM-type phase,
resulting from an (appropriate) effective neutrino mass matrix and the
charged leptons mass matrix, can be zero, while the (total) high energy
theory violates CP\cite{gui}. In order to see this, take, in the
parametrization of Eqs. (\ref{only}, \ref{odoe}) for $M_D$ and $M_e$, all
1-2 angles zero. This will guarantee a zero CKM-type phase, for the
effective low energy theory. In addition, for this simple case, we construct
a class of matrices (resulting in $I_i\neq 0$ but) with no CP violation
Majorana phases for the approximate effective low energy theory. This class
depends on two parameters $\theta $, $\alpha $: 
\begin{equation}
\label{canbe}
\begin{array}{l}
M_D=D\ F^TK_\beta \\ 
M_e=K_\alpha F^TD_e\ FK_\alpha ^{*} \\ 
M_R=D_R 
\end{array}
\ ;~F=\left[ 
\begin{array}{lll}
1 & 0 & 0 \\ 
0 & \frac 1{\sqrt{2}} & \frac 1{
\sqrt{2}} \\ 0 & \frac{-1}{\sqrt{2}} & \frac 1{\sqrt{2}} 
\end{array}
\right] \ 
\end{equation}
where $D=m\ {\rm {diag}[\lambda ,\sin (\theta ),\cos (\theta )]}$, $D_R=M\ 
{\rm {diag}[\mu ,b^{+},\ b^{-}]}$, with $b^{\pm }=(1\pm \sin (2\theta )\cos
(\alpha ))$ and $\lambda $, $\mu $, $m$, $M$ real. $K_\alpha ={\rm {diag}%
[1,1,e^{i\alpha }]}$, $K_\beta ={\rm {diag}[1,1,ie^{i\beta }]}$ with $\tan
(\beta )=-\tan (2\theta )\sin (\alpha )$. $D_e$ may be any real diagonal
matrix. For this specific choice of parametrization for the neutrino Dirac,
Majorana and charged lepton mass matrices and in a weak-basis where the
charged lepton mass matrix is diagonal, the low energy effective neutrino
mass matrix has the special form, 
\begin{equation}
\label{special}m_{eff}=-\widehat{M}_D\ D_R^{-1}\ \widehat{M}_D^T=\left[ 
\begin{array}{ccc}
r & 0 & 0 \\ 
0 & 0 & c \\ 
0 & c & c^{\prime } 
\end{array}
\right] 
\end{equation}
where $\widehat{M}_D=FK_\alpha D\ F^TK_\beta $ and where $r$ is real and the 
$c$'s are complex. As a result (note that $(m_{eff})_{22}=0$) all the phases
in the $c$'s are irrelevant and can be absorbed through (or into) the
righthanded charged lepton fields. Thus, not only the CKM-type phase but
also the Majorana phases are zero and there is no form of CP-violation in
the low energy theory. However, for $M_D^{\dagger }M_D$ we obtain 
\begin{equation}
\label{mdmd}M_D^{\dagger }M_D=\frac{m^2}2\left[ 
\begin{array}{lll}
2\lambda ^2 & 0 & 0 \\ 
0 & 1 & \cos (2\theta )e^{i\beta } \\ 
0 & \cos (2\theta )e^{-i\beta } & 1 
\end{array}
\right] 
\end{equation}
where, remember, $\tan (\beta )=-\tan (2\theta )\sin (\alpha )$. Therefore,
in contrast with the approximate effective low energy theory, CP-violation
does indeed occur for the (total) high energy theory and the CP violation
leptogenisis invariants $I_i$, computed with Eq. (\ref{week}), are different
from zero: 
\begin{equation}
\label{finaliii}
\begin{array}{l}
I_1=\ 
\frac{m^4M^4}2\frac{\sin ^2(2\theta )\cos ^2(\alpha )-1}{\tan ^2(2\theta
)\sin ^2(\alpha )+1}\sin (2\theta )\sin (4\theta )\sin (2\alpha ) \\  \\ 
I_2=m^4M^6 
\frac{\sin ^4(2\theta )\cos ^4(\alpha )-1}{\tan ^2(2\theta )\sin ^2(\alpha
)+1}\sin (2\theta )\sin (4\theta )\sin (2\alpha ) \\  \\ 
I_3=\frac{m^4M^8}2\frac{\left( \sin ^2(2\theta )\cos ^2(\alpha )-1\right) }{%
\tan ^2(2\theta )\sin ^2(\alpha )+1}^3\sin (2\theta )\sin (4\theta )\sin
(2\alpha ) 
\end{array}
\ 
\end{equation}

{\bf Conclusions}

Using RCP we reduce, significantly, the number of CP violating complex
phases in the Yukawa parameters in two important cases. In the case of an
(effective) $3\times 3$ Majorana neutrino mass matrix, we obtain only one
complex CP violating phase, in contrast with the general structure, where
there is one CKM-type and 2 Majorana type complex phases. For the See Saw
extension of the SM with three righthanded neutrinos, and in connection with
CP violation in leptogenis, we reduce the usual 6 complex phases to only 2.
We also show that RCP is usefull in the analys of some specific scenarios.
In particular, we give an example where, in the case of the see saw
mechanism, the (total) high energy theory violates CP even when all forms
CP-violation are absent in the approximate effective low energy theory.

\end{document}